\documentclass[aps,prl,
groupedaddress,amssymb,
showpacs,nofootinbib]{revtex4}

\usepackage{amsmath,amssymb}

\usepackage[usenames]{color}

\usepackage{graphicx}
\usepackage{mathptmx}
\usepackage{bbm}
\usepackage{bm}
\usepackage{times}

\newcommand{\beq}{\begin{eqnarray}}
\newcommand{\eeq}{\end{eqnarray}}

\begin{document}

\onecolumngrid



\title{Seeing asymptotic freedom in an exact correlator of a large-$N$ matrix field theory}



\author{Peter \surname{Orland}}

\email{orland@nbi.dk}


\affiliation{1. Baruch College, The 
City University of New York, 17 Lexington Avenue, 
New 
York, NY 10010, U.S.A. }

\affiliation{2. The Graduate School and University Center, The City University of New York, 365 Fifth Avenue,
New York, NY 10016, U.S.A.}

\begin{abstract}

Exact expressions for correlation functions are known for the large-$N$ (planar) limit of the 
$(1+1)$-dimensional ${\rm SU}(N)\times {\rm SU}(N)$ principal chiral sigma model. These were obtained
with the form-factor bootstrap, an entirely nonperturbative method. The large-$N$ solution of this asymptotically-free 
model is far less trivial than
that of  O($N$) sigma model (or other isovector models). Here 
we study the Euclidean two-point correlation function
$N^{-1}\langle {\rm Tr}\,\Phi(0)^{\dagger} \Phi(x)\rangle$, where $\Phi(x)\sim Z^{-1/2}U(x)$ is the scaling field and
$U(x)\in {\rm SU}(N)$ is the bare field. We express the two-point function
in terms of
the spectrum of the operator $\sqrt{-d^{2}/du^{2}}$, where $u\in (-1,1)$. At short distances, this expression
perfectly matches the result from the perturbative renormalization group.

\end{abstract}


\maketitle

Green's functions of quantum chromodynamics (QCD) cannot be calculated at large separations 
analytically. Currently, only numerical lattice
calculations suffice for this purpose. On the other hand, perturbation theory
can be used to understand short-distance behavior in any asymptotically-free theory, such as QCD. In lower dimensions, there are field-theoretic 
models with asymptotic freedom, which can be studied mathematically. A nontrivial example is the principal
chiral sigma model (PCSM) of a matrix field $U(x) \in {\rm SU}(N)$, $N\ge 2$, where $x^{0}$ and 
$x^{1}$ are the time and space coordinates, respectively. Here, the large-$N$ limit of the PCSM will be considered. The PCSM is a matrix model, not an 
isovector model (such as the ${\rm O}(N)$ or ${\mathbb CP}(N-1)$ sigma models or the Gross-Neveu model). The PCSM's large-$N$ limit has not been solved by saddle-point methods. Its Feynman diagrams are 
truly planar, not linear. Finally, the PCSM has nontrivial field renormalization, even in the large-$N$ limit; this means that its correlation functions are not those of a free
field theory, in this limit. In all of these respects, the PCSM resembles QCD substantially better than isovector field theories.

In this letter, an exact expression for a correlation function of the large-$N$ PCSM is studied at short distances, where it is found to obey a power-law decay law. At large distances, this
correlation function has exponential decay. Thus, the solution clearly illustrates both ultraviolet freedom and an infrared mass gap. Furthermore, the ultraviolet behavior of this
nonperturbatively-obtained correlation function has precisely the behavior expected from the perturbative renormalization group. The key to the short-distance behavior
is the spectrum of an interesting integro-differential operator on functions of the open interval $(-1,1)$.

The PCSM has the action
\begin{eqnarray}
S=\frac{N}{2g_{0}^{2}}\int d^{2}x \;\eta^{\mu\nu}\;{\rm Tr}\,\partial_{\mu}U(x)^{\dagger}\partial_{\nu}
U(x),
\label{action}
\end{eqnarray}
where $\mu, \nu=0,1$, $\eta^{00}=1$, $\eta^{11}=-1$, $\eta^{01}=\eta^{10}=0$, where $g_{0}$ is the coupling (which is held fixed as $N\rightarrow\infty$). This action is invariant under the global transformation 
$U(x)\rightarrow V_{L}U(x)V_{R}$,  for two constant matrices $V_{L}, \,V_{R}\in {\rm SU}(N)$. The renormalized field operator $\Phi(x)$ is an average of $U(x)$ over a region of size $b$, where $\Lambda^{-1}<b\ll m^{-1}$, where $\Lambda$ is an ultraviolet cutoff and $m$ is the mass of the fundamental excitation. 

For matrix models in more than one dimension, there is no general approach to 
summing the planar diagrams. The 
PCSM, however, has the virtue of being integrable. Integrability 
is not sufficient to determine Green's functions, although the S matrix has been known for three decades \cite{Smatrix}. Recently, both 
integrability and the $1/N$-expansion were 
combined
to find the $N\rightarrow \infty$ limit of Green's functions \cite{PO}, \cite{ACC}. This was done using Smirnov's axioms for form factors \cite{Smirnov}. The
form-factor bootstrap method has a long history \cite{EFF}. A detailed comparison of the 
$1/N$-expansion and form factors of the
O(N) sigma model is in Ref. \cite{BKF}.

In this letter, we study an exact non-perturbative expression for the two-point function of the scaling field $\Phi(x)$,
found in the second of Ref. \cite{PO}. The scaling field $\Phi$ is normalized by 
$\langle 0\vert \Phi(0)_{b_{0} a_{0}}\vert P,\theta, a_{1}, b_{1}\rangle
=N^{-1/2}\delta_{a_{0} a_{1}}\delta_{b_{0} b_{1}}, \label{norm}$
where the ket on the right is a one particle ($r=1$) state, with rapidity $\theta$. This field is a complex $N\times N$ matrix, which is
not directly proportional to the unitary matrix $U(x)$. Nonetheless we write
$\Phi(x) \sim {\rm Z}(g_{0},\Lambda)^{-1/2}U(x)$,
which means that (the time-ordering is optional)
\begin{eqnarray}
\frac{1}{N}\left \langle 0\vert {\rm Tr}\;\Phi(x) \Phi(0)^{\dagger} \vert 0\right\rangle
={\rm Z}(g_{0},\Lambda)^{-1}
\frac{1}{N}\left \langle 0\vert {\rm Tr}\;U(x)U(0)^{\dagger} \vert 0\right\rangle.
\label{renormalization}
\end{eqnarray}
It would be interesting to know the relation of
the scaling field used in lattice simulations \cite{NNV} to that defined above, which
is not yet clear to the author. Particle masses are given by the sine formula:
$m_{r}=m\;\sin(\pi N^{-1}r)/\sin(\pi N^{-1}),\; \; r=1,\dots,N-1$,
but in the large-$N$ limit, only the $r=1,$ $r=N-1$ states (the elementary 
particle and antiparticle) survive. The binding energies of the other states vanish. The residues
of their poles in S-matrix elements also vanish.

The renormalization factor 
${\rm Z}(g_{0}(\Lambda),\Lambda)$ vanishes as $\Lambda\rightarrow \infty$ and the coupling
$g_{0}(\Lambda)$ runs so that the mass gap
$m(g_{0}(\Lambda), \Lambda)$ is independent of $\Lambda$.  For $m\vert x\vert\gg 1$, the expression (\ref{renormalization}) decays exponentially, as expected. We find that for 
$m\vert x\vert\ll 1$, the time-ordered product of two scaling field operators behaves as
\begin{eqnarray}
\frac{1}{N}\left \langle 0\vert \; {\mathcal T}\;{\rm Tr}\;\Phi(x) \Phi(0)^{\dagger} \vert 0\right\rangle 
=C_{2}(\ln m\vert x\vert)^2 +C_{1}\ln m\vert x\vert+C_{0}+O(1/\ln m\vert x\vert)
, \label{asymptform}
\end{eqnarray}
for some constants $C_{2}$, $C_{1}$, etc. The leading term is exactly what a perturbative-renormalization-group analysis 
implies. We consider this to be a striking validation of the form-factor bootstrap.

Let us recall the argument for (\ref{asymptform}) (see for example, Ref. \cite{PB}). For convenience, we perform the Wick rotation $x^{0}\rightarrow {\rm i}x^{0}$, to
obtain the regularized Euclidean correlation function
$G(\vert x \vert, \Lambda)=N^{-1}\left \langle 0\vert \;{\mathcal T}\;{\rm Tr}\;\Phi(x) \Phi(0)^{\dagger} \vert 0\right\rangle$. This function and the coupling
$g_{0}(\Lambda)$ satisfy the renormalization group equations
\begin{eqnarray}
\frac{\partial\ln G(R, \Lambda)}{\partial\ln \Lambda}= \gamma(g_{0})=\gamma_{\,1}g_{0}^{2}+\cdots\;,\;\;\;
\frac{\partial g_{0}^{2}(\Lambda)}{\partial \ln \Lambda}=\beta(g_{0}^{2})=-\beta_{1}g_{0}^{4}+\cdots, \label{RGE}
\end{eqnarray}
respectively. The coefficients of the anomalous dimension $\gamma(g_{0})$ and the beta function $\beta(g_{0})$ are $\gamma_{\;1}=(N^{2}-1)/(2\pi N^{2})$ and
$\beta_{1}=1/(4\pi)$. For large $\Lambda$, $G(R,\Lambda)$ becomes a function of the product of the two variables $G(R\Lambda)$. Integrating (\ref{RGE})
yields the leading behavior
\begin{eqnarray}
G(R,\Lambda) \sim C[\ln (R\Lambda)]^{\gamma_{\,1}/\beta_{1}}\;. \label{asympt}
\end{eqnarray}
As $N\rightarrow \infty$, the power $\gamma_{\,1}/\beta_{1}$ approaches $2$.

The exact Wightman function (in Minkowski spacetime) of the 
product of two fields (that is, with no time-ordering) is ${\mathcal W}(x)= N^{-1}\langle 0\vert  {\rm Tr}\;\Phi(x) \Phi(0)^{\dagger} \vert 0\rangle$. This function is
\cite{PO}
\begin{eqnarray}
{\mathcal W}(x)
=\int_{-\infty}^{\infty} \frac{d\theta_{1}}{4\pi} e^{{\rm i}p_{1}\cdot x}+ \frac{1}{4\pi}\sum_{l=1}^{\infty} \int_{-\infty}^{\infty} d\theta_{1}\cdots \int_{-\infty}^{\infty} d\theta_{2l+1}
e^{{\rm i}\sum_{j=1}^{2l+1}p_{j}\cdot x}\;
\prod_{j=1}^{2l}\frac{1}{(\theta_{j}-\theta_{j+1})^{2}+\pi^{2}}
\;,
\label{series}
\end{eqnarray}
where $\theta_{j}$ are rapidities and $p_{j}=m(\cosh\theta_{j}, \sinh\theta_{j})$ are the corresponding momentum vectors, for $j=1,\dots 2l+1$. The right-hand side
of (\ref{series}) is difficult to evaluate. For spacelike separation $x^{0}=0$, it decays exponentially with $\vert x^{1} \vert$. Our purpose here to evaluate (\ref{series})
for small timelike separation $x^{1}=0$, $x^{0}\ll m^{-1}$. For this case, the Wightman function is equal to the time-ordered expectation value on
the left-hand side of (\ref{asymptform}). Eq. (\ref{series}) or an approximation to it has not yet been obtained in any program 
to solve the large-$N$ PCSM directly from the action (\ref{action}). Perhaps, one day, this will be done (a recent proposal is in Ref. \cite{resurg}).

To study the two-point function at short distances, it is convenient to Wick-rotate the time variable to Euclidean space as above. Setting $x^{1}=0$ and
replacing $x^{0}$ by ${\rm i}R$, $R>0$, changes the phases in (\ref{series}) by
$\exp{{\rm i}p_{j}\cdot x} \rightarrow \exp{-mR\cosh \theta_{j}}$. We define $L=\ln \frac{1}{mR}$.
As $mR$ becomes small, $\exp{-mR\cosh \theta_{j}}$ becomes approximately the characteristic
function of $(-L,L)$, equal to unity for $-L<\theta<L$ and zero everywhere else. This 
is mathematically similar to the formation of walls in the Feynman-Wilson gas \cite{FW}. This trick
was used to find the scaling behavior of Ising-model correlation functions \cite{CMYZ} from the exact form
factors \cite{Ising}. The short-distance Euclidean two-point function is
\begin{eqnarray}
G(mR)
=\frac{L}{2\pi} + \frac{1}{4\pi}\sum_{l=1}^{\infty} \int_{-L}^{L} d\theta_{1}\cdots \int_{-L}^{L} d\theta_{2l+1}\;
\prod_{j=1}^{2l}\frac{1}{(\theta_{j}-\theta_{j+1})^{2}+\pi^{2}}
\;.
\label{Gseries}
\end{eqnarray}
Notice that the first term of (\ref{series}), which is the Wightman function of a free massive field, corresponds to the first term of
(\ref{Gseries}) which is the Euclidean correlation function of a massless field. The expression (\ref{Gseries}) is the partition function
of a polymer in a box of size $2L$. The $j^{\rm th}$ atom in the polymer chain is located at $\theta_{j}$. There is a long-range potential energy
$\ln [(\theta_{j}-\theta_{j+1})^2+\pi^2]$, between atoms connected on the chain.

It is convenient to rescale the integration variables by $\theta_{j}=Lu_{j}$, so that (\ref{Gseries}) becomes
\begin{eqnarray}
G(mR)
=\frac{L}{2\pi} + \frac{L}{4\pi}\sum_{l=1}^{\infty} \int_{-1}^{1} du_{1}\cdots \int_{-1}^{1} du_{2l+1}\;
\prod_{j=1}^{2l}\frac{1}{L[(u_{j}-u_{j+1})^{2}+(\pi/L)^{2}]}
\;.
\label{Gseries1}
\end{eqnarray}
There is a close relation between the terms of  (\ref{Gseries1}) and the fractional-power-Laplace operator $\Delta^{1/2}=\sqrt{-d^{2}/du^{2}}$. The spectrum of
$\Delta^{\alpha/2}$, with real $\alpha\in (0,2)$, is
a subject of active mathematical investigation  \cite{JE}, \cite{FracLap}. The self-adjoint extension of the operator $\Delta^{1/2}$ on $u\in (-1,1)$ has an infinite
set of discrete eigenvalues $\lambda_{n}$, of the eigenfunctions $\varphi_{n}(u)$, $n=1,2,\dots$, with $0<\lambda_{1}<\lambda_{2}<\cdots$, with
$\varphi_{n}(\pm 1)=0$. Another polymer statistical system in which a fractional power of the second derivative plays a role is described in Ref. \cite{RP}.

Here is a quick introduction to the operator $\Delta^{1/2}$, {\em via} the Poisson semigroup. Let us forget 
the restriction to the open interval and extend the rapidity variables to the real line $(-\infty,\infty)$. Consider the transfer operators $P(a)$, whose
matrix elements are defined by $\langle u^{\prime} \vert P(a) \vert u\rangle =a[(u^{\prime}-u)^{2}+a^{2}]^{-1}\pi^{-1}$, where $u^{\prime}$
and $u$ are arbitrary real numbers. These operators form the Poisson semigroup \cite{JE}, with the composition law 
$P(a)P(b)=P(a+b)$. Specifically, $P(a)=\exp -a\Delta^{1/2}$, where $\Delta^{1/2}=\sqrt{-d^{2}/du^{2}}$. 

Explicitly, the square root of the Laplacian on a function $f(u)$, vanishing for $u\notin (-1,1)$, is \cite{JE}
\begin{eqnarray}
{\Delta}^{1/2}f(u)=\frac{1}{\pi}\int_{-1}^{1}du^{\prime}\;{\rm PV}\;\frac{f(u^{\prime})-f(u)}{(u^{\prime}-u)^{2}}\;, \label{FL}
\end{eqnarray}
where PV denotes the principal value. This operator has an infinite
set of discrete eigenvalues $\lambda_{n}$, of the eigenfunctions $\varphi_{n}(u)$, $\Delta^{1/2}\varphi_{n}=\lambda_{n}\varphi_{n}$, $n=1,2,\dots$, with $0<\lambda_{1}<\lambda_{2}<\cdots$, with
$\varphi_{n}(\pm 1)=0$. Now for
$u, u^{\prime}\in (-1,1)$, we define the operator $H(L)$ by
\begin{eqnarray}
\frac{1}{L[(u-u^{\prime})^{2}+(\pi/L)^{2}]}
=
\langle u^{\prime} \vert e^{-\frac{\pi}{L}H(L)}
\vert u \rangle. \label{TM}
\end{eqnarray}
Then (\ref{FL}), (\ref{TM}) and a straightforward calculation show that $H(L)$ is an approximation to $\Delta^{1/2}$, {\em i.e.}, $H(L)={\Delta}^{1/2}+O(1/L)$, with spectrum
\begin{eqnarray}
H(L)\varphi_{n}(u,L)=\lambda_{n}(L)\varphi_{n}(u,L), \; \int_{-1}^{1}du\,\vert\phi_{n}(u,L)\vert^{2}=1,\;\lambda_{n}(L)=\lambda_{n}+O(1/L), \;\varphi_{n}(u,L)=\varphi_{n}(u)
+O(1/L)\;.
\label{pertspec}
\end{eqnarray}

Summing 
over $l$ in
Eq. (\ref{Gseries1}) yields, from (\ref{pertspec}),
\begin{eqnarray}
G(mR)=\frac{L}{4\pi} \int_{-1}^{1}du^{\prime} \int_{-1}^{1} du \; 
\langle \,u^{\prime}\; \vert \;\frac{1}{1-e^{-2\pi H(L)/L}}\; \vert \,\,u \,\rangle 
=\frac{L}{4\pi}\sum_{n=1}^{\infty} \left\vert  \int_{-1}^{1}du\; \varphi_{n}(u,L)      \right\vert^{2} \frac{1}{1-e^{-2\pi\lambda_{n}/L+O(1/L^{2})}}. \label{resum}
\end{eqnarray}
Parenthetically, we note that $\int_{-1}^{1}du\, \phi_{n}(u,L)=0$ for even $n$. We 
split (\ref{resum}) into two sums:
\begin{eqnarray}
G(mR)=\frac{L}{4\pi}\sum_{\lambda_{n}\le L/2\pi} \left\vert  \int_{-1}^{1}du\; \varphi_{n}(u,L)      \right\vert^{2} \frac{1}{1-e^{-2\pi\lambda_{n}/L+O(1/L^{2})}}
+\frac{L}{4\pi}\sum_{\lambda_{n}>L/2\pi} \left\vert  \int_{-1}^{1}du\; \varphi_{n}(u,L)      \right\vert^{2} \frac{1}{1-e^{-2\pi\lambda_{n}/L+O(1/L^{2})}}\;.
\label{splitresum}
\end{eqnarray}
The second term in (\ref{splitresum}) cannot diverge as $L\rightarrow \infty$, hence gives no contribution to either $C_{1}$ or $C_{2}$. For
\begin{eqnarray}
\frac{L}{4\pi}\sum_{\lambda_{n}>L/2\pi} \left\vert  \int_{-1}^{1}du\; \varphi_{n}(u,L)      \right\vert^{2} \frac{1}{1-e^{-2\pi\lambda_{n}/L+O(1/L^{2})}}\; \lesssim \;
\frac{L}{4\pi}\sum_{\lambda_{n}>L/2\pi} \left\vert  \int_{-1}^{1}du\; \varphi_{n}(u,L)    \right\vert^{2}  \frac{1}{1-e^{-1}} \;, \nonumber
\end{eqnarray}
and the sum over $n$ on the right-hand side is roughly 
\begin{eqnarray}
\sum_{\lambda_{n}>L/2\pi} \left\vert  \int_{-1}^{1}du\; \varphi_{n}(u,L)    \right\vert^{2} \sim \frac{1}{L}\;. \nonumber
\end{eqnarray}
The first
term in (\ref{splitresum}) may be expanded in powers of $1/L$ to yield
\begin{eqnarray}
\frac{L}{4\pi}\sum_{\lambda_{n}\le L/2\pi} \left\vert  \int_{-1}^{1}du\; \varphi_{n}(u,L)      \right\vert^{2} \frac{1}{1-e^{-2\pi\lambda_{n}/L+O(1/L^{2})}}
\;=\;\frac{L}{4\pi}\sum_{\lambda_{n}\le L/2\pi} \left\vert  \int_{-1}^{1}du\; \varphi_{n}(u)      \right\vert^{2} 
\frac{L}{2\pi\lambda_{n}}+O(L). \nonumber
\end{eqnarray}
Extending the  sum over $n$ from zero to infinity gives the leading coefficient in (\ref{asymptform}):
\begin{eqnarray}
C_{2}=\frac{1}{8\pi^{2}}\sum_{n=1}^{\infty}\left\vert  \int_{-1}^{1}du\; \varphi_{n}(u)      \right\vert^{2} \lambda_{n}^{-1}.\label{C12}
\end{eqnarray}
An upper bound on the leading 
coefficient $C_{2}$ is obtained by replacing $\lambda_{n}$ in (\ref{C12}) by $\lambda_{1}$, and using completeness: 
$\sum_{n} \vert \int_{-1}^{1} du\,\phi_{n}(u)\vert^2=2$. Thus
$C_{2}<\frac{1}{4\pi^{2}\lambda_{1}}=0.0219$,
from the best known value of $\lambda_{1}=1.1577$, found in the second and third of
Refs. \cite{FracLap}. It is
interesting that without much detailed knowledge 
of the properties of $H(L)$ or of the square root of the Laplacian, we have established
the ultraviolet behavior (\ref{asymptform}) of the two-point correlation function. An evaluation of $C_{1}$ would require a better understanding of the spectrum
of $H(L)$.

To conclude, we believe the correlators of ${\rm SU}(\infty)\times {\rm SU}(\infty)$ PCSM are now
understood almost as well as those of the Ising model \cite{Ising}. The exact $N\rightarrow \infty$ correlation function argued for in Ref. \cite{PO} displays
massive behavior at large distances. We have found precisely the short-distance behavior predicted with the 
perturbative beta function and anomalous dimension. This strengthens our confidence in
the form factors \cite{PO}, \cite{ACC}, which led to this result. 

It is a 
pleasure to thank Dr. Axel Cort\'es Cubero, whose suggestions led to the expression (\ref{Gseries1}), and Dr. Timothy Budd for discussions about the Poisson semigroup and 
the fractional-power Laplacian. This work was supported in 
part by a grant from the PSC-CUNY. 

\end{document}